\documentstyle[multicol,prl,aps,epsf,psfig,epsfig]{revtex}

\begin{document}

\title
{
Clustering properties of a generalised critical Euclidean network
}
\author
{
Parongama Sen$^1$ and S. S. Manna$^2$
}
\address
{
$^1$Department of Physics, University of Calcutta,
    92 Acharya Prafulla Chandra Road, Kolkata 700009, India. \\
$^2$Satyendra Nath Bose National Centre for Basic Sciences
    Block-JD, Sector-III, Salt Lake, Kolkata-700098, India \\
}
\maketitle
\begin{abstract}
Many real-world networks exhibit scale-free feature, have a small diameter and
a high clustering tendency.
We have studied the properties of  a growing network, 
which has all these features, 
in which  an
incoming node is connected to its $i$th predecessor of degree $k_i$ 
with a  link of length $\ell$ using a probability proportional
to  $k^\beta _i \ell^{\alpha}$. 
For $\alpha > -0.5$,
the network is scale free at $\beta = 1$ 
with the degree distribution $P(k) \propto k^{-\gamma}$ and $\gamma = 3.0$ 
as in the Barab\'asi-Albert model ($\alpha =0, \beta =1$).
We find a phase boundary in the $\alpha-\beta$ plane along which the network is
scale-free. 
Interestingly,
we find scale-free behaviour  even for $\beta > 1$ for
$\alpha < -0.5$ where the existence of a new universality class is indicated
from  the behaviour of the degree distribution and the clustering coefficients.  
The  network has a small diameter in the entire
scale-free  region. 
The clustering coefficients emulate the behaviour of
most real networks for increasing  negative values of
$\alpha$ on the phase boundary. 


\end{abstract}

\begin{multicols}{2}
Recent  studies of  many complex real-world networks of diverse nature, e.g.,  social 
networks, biological networks, electronic communication networks 
  etc. reveal some striking similarities in their
underlying structures \cite{review}. The diameter ${\cal D}$, a measure of 
the  topological extension 
of the network, the clustering coefficient ${\cal C}$, a measure of the local correlations 
among the links of the network and the nodal degree distribution $P(k)$ are some of the few 
important quantities 
which exhibit the similarities among the different networks.
Many of these
networks exhibit 
   small-world  network (SWN)-like properties 
\cite{watts}, i.e.,  the diameter ${\cal D}(N)$ of the network scales 
logarithmically 
with the number of nodes $N$ while the clustering coefficient has a high value.  
In some of these  networks,
there is no characteristic scale manifested by the typical 
power law decay of the tail of the degree distribution: $ P(k) \propto k^{-\gamma}$
  \cite{BA},
  where  $ P(k)$ is the number of nodes which are linked with  $k$ other nodes.
  These networks are called scale-free networks (SFN).
 
Typically, the clustering coefficient measures the 
conditional probability that an arbitrary pair of nodes are linked, provided
both are linked to a third node. 
The clustering coefficient can be studied as a function of two different 
variables: ${\cal C}(N)$, the clustering coefficient per node averaged 
over all $N$ nodes as a function of the network size $N$ and ${\cal C}(k)$, 
the clustering coefficient per node averaged over all nodes with degree $k$ 
as a function of $k$. Obviously ${\cal C} (N) = \Sigma_k P(k) {\cal C} (k)/\Sigma_k P(k)$.

In some recent studies
  \cite{RB,ves}, it was shown that several real networks, like the actor network, 
  language network,  the Internet at the autonomous system level  etc.,  
which are known to exhibit  scale-free behaviour and have  small diameters, 
have another common feature, i.e., 
${\cal C}(k) $ has a power law dependence: ${\cal C}(k) \propto k^{-1}$
whereas the  total clustering coeffcient  ${\cal C}(N)$  has a high value.

 Attempts to capture  the  three features of  
 small diameter, high clustering and absence of a characteristic scale,
 which occur in 
 many real world networks, in a single model,  have been
 faced with certain difficulties. 
 The first model to mimic a small-world network is the  Watts-Strogatz model 
 (WS) \cite{watts}. 
 Here  the nodes are arranged on a ring 
 with links to
 the nearest neighbours and  small-world
 features can be achieved  by re-wiring the nearest neighbour bonds to 
 randomly link
    an arbitrary  pair of nodes even with a very small probability.
 However the nodal degree distribution 
 in the WS model failed to show  scale-free feature. The Barab\'asi-Albert (BA) model is a  prototype for a 
 SFN in which  the network is grown by adding nodes one by one, and
 a new  node gets attached to an older one with a probability
 proportional to its degree.
 Although the scale-free property was successfully achieved and the network
 had a small diameter, the clustering coefficient   ${\cal C}(N)$  
 showed a power law decay with $N$ (${\cal C}(N) \propto N^{-0.75}$), while ${\cal C}(k)$ remained
 a constant with $k$ \cite{review,RB}, thus failing to capture the feature of high clustering tendency  of real networks.

 Successful attempts to capture all the desirable features
 of a network have been made by defining other models  
 \cite{RB,Doro,klemm,holme,jost,szabo} subsequently.
 For example, in a deterministic growing graph \cite{Doro}, which is argued to  simulate
 a citation network,
 exact calculations showed that it has small diameter, scale-free
 feature as well as ${\cal C}(k) \propto  1/k$. 
 In \cite{holme,szabo}, suitable modifications are 
 done to generate triads (and consequently a high clustering coeffcient) 
 in an otherwise
 BA type of growing network.
 In \cite{klemm},  an old node is deactivated with a probability
 proportional to its inverse degree in a growing network to get
 a high clustering coefficient.
 In \cite{jost}, spatial distances have been incorporated
 in some specified manner which also gave the desired features
 of a real network to a large extent.
A power law dependence of  ${\cal C}(k) $ can be obtained in deterministic 
and stochastic scale-free networks 
with hierarchical structure also
\cite{RB}.

While in a majority  of  real-world  scale-free networks  
 ${\cal C}(k) \propto 1/k$ , some other  networks like the Internet router network,
the power grid network \cite{RB,ves} of the Western United states and
the Indian railway network \cite{train} (which does not have  
 scale-free behaviour)
   showed a different behaviour: 
 ${\cal C}(k)$ shows no dependence or logarithmic dependence on $k$. 
  In \cite{RB}, this  behaviour  was  
  argued to be due to   the presence of  geographical organisation in   
  such  networks in the sense that there  are actual 
  physical connection between the
  nodes and the networks are defined in real space. 
 A comparison of the clustering coefficients in a model network with and without 
 geographical organisation  could therefore help to
 understand the relation between geographical organisation and clustering better.  
  It should be pointed out here  that 
  ${\cal C}(k)$ is also a constant in the BA model where 
   a metric is  not defined.

 In a network defined in real space, the spatial distance 
 between the nodes is expected to play an important role
 in constructing the links. On the other hand, the rule
 of preferential attachment has been very successful in achieving 
 the scale-free feature and small diameter of a network.
We  have  therefore considered a  growing network in which 
both the preferential attachment  
as well as the spatial
  distances are  parametrically incorporated in the attachment
  probability and 
   can be independently tuned. 
Here we would like to mention that although some networks are
not defined in real space, spatial distances
are still expected to be implicitly involved, e.g., in a social network, 
people in the same locality
are much more likely to know and influence each other.
Although the concept of geographical locality  does not exist
explicitly in all networks,  one can still define 
a ``closeness'' factor in many networks, e.g., in the citation network,
a paper is likely to be cited with a higher probability  when the contents of it is ``close''
to that of the citing paper.
  
  The network  we have under consideration 
  is  evolved from the time $t=0$  and at time $t$   an 
  incoming node gets attached
  to the $i$-th  node with degree $k_i$, at a distance $\ell $, according to the
  following probability:
\begin{equation}
\label{eq1}
  \pi_i(t) \sim k^{\beta}_i(t) \ell^{\alpha}.
\end {equation}

\begin{center}
\begin{figure}
\noindent \includegraphics[clip,width= 5cm]{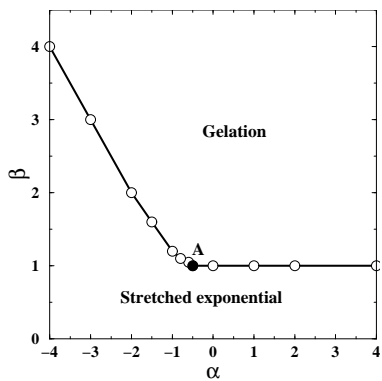}
\caption{The phase diagram of the network in the $\alpha-\beta$ plane.
Scale-free behaviour is observed only at the boundary. The point A indicates 
 a change in the critical behaviour: to the right of A the
critical behaviour is of the BA type while to its left we find a new
critical behaviour.}
\end{figure}
\end{center}


In some earlier studies \cite{jost,yook,mannaps,xulvi,barth},  spatial dependence 
in a growing network has been
studied   
where the attachment probability is dependent on the spatial distances between the node.
In \cite{jost}, the clustering coefficients were also calculated.
However, the spatial dependence was not incorporated in a way it could be 
systemetically studied.  
In the present study, the aims are   (a) to identify the regions where the network 
is scale free in the $ \alpha - \beta$ plane and (b) to study the behaviour of the clustering coefficient as a function of
the parameters $\alpha $ and $\beta$ and (c) to check 
whether the  diameter of the network  and 
the average shortest distances 
scale logarithmically with the number of nodes.

In this network, each incoming node gets bonded to $m$  distinct nodes. 
In order to study clustering
properties $m$ should be at least equal to two
($m = 1$ would lead to a tree like structure
with no loops and all clustering coefficients are trivially zero here.)
Results for some  limiting cases of the model defined by (1) are known.  
The $\alpha = 0$ and $\beta = 1$ case corresponds to the 
 scale-free BA network  \cite{BA}. Networks with  $\alpha = 0$ and arbitrary values of $\beta$,
  considered in \cite{Red},  showed that scale-free
 behaviour existed only for $\beta = 1$. For $\beta > 1$, there
 is a tendency of the incoming nodes to get connected to a single node and
 this behaviour is termed ``gelation''. For $\beta < 1.0$, the
 behaviour of the degree distribution is stretched exponential. 
 The effect of Euclidean
 distances were incorporated in a BA kind of network  \cite{mannaps,xulvi}
 by keeping $\alpha $ non-zero and $\beta = 1$ where the network is defined in
 a $d$-dimensional Euclidean plane.
 It was found  that the scale free behaviour persists above a certain critical 
 value of $\alpha$ which depends on the spatial dimensionality. 
 Below this value of $\alpha$, the stretched exponential behaviour of the 
 degree distribution was again observed.

 
 We considered  a one-dimensional space with periodic boundary condition 
 where the nodes 
occupy   the position  $x$   with  $0 < x \leq 1$.
Initially we have $m_0$ nodes connected to each other.
 First we  investigate the scale-free properties of the
  model by studying the degree distribution $P(k)$.
 We vary both $\alpha$ and $\beta$ 
 and observe the behaviour of $P(k)$  to obtain a phase diagram. 
 Results for $m_0 =m = 1$ and $m_0 = m =3$ 
 showed that  the critical behaviour is independent  of the value of $m$ as in the BA model.
 We noted  several interesting features:

 1. In the $\alpha-\beta $ plane, there exists a phase boundary along which the network is scale-free.
 Above this boundary it shows a gelation-like  behaviour as in  \cite{Red}.
 Below this boundary the degree distribution is stretched exponential as 
 was observed in
 \cite{mannaps},  \cite{xulvi} and \cite{Red}.

 2. Scale free behaviour is observed to occur  at the critical value $\beta_c = 1$  for all values of
 $\alpha \ge  -0.5$. 
 For lower values of $\alpha$ it occurs at higher values of $\beta$.
  For values of $\alpha < - 2.0$ the phase boundary is linear given by the 
  equation $\alpha_c + \beta_c = 0$.

 3. Although the scale-free property is observed along the entire phase
 boundary, there is a difference in the behaviour of the degree distribution 
 $P(k)$. 
 While $P(k) \sim k^{-\gamma }$ everywhere, $\gamma \sim  2.7$ for 
$\alpha   < -0.5$
 and $\gamma = 3.0$ (as in the BA model) for $\alpha >  -0.5$.

 The phase diagram is shown in   Fig. 1.
We would like to emphasise two points from the above
observations. First, even though the case $\beta \ne 1$ has been studied 
earlier \cite{Red}, the only point at which scale-free behaviour was observed was
at $\beta_c = 1$ while here one can get scale-free behaviour even at
$\beta_c > 1$ by tuning the distance dependence factor. 
Secondly,  the exponent $\gamma = 2.7 \pm 0.1 $ for $\alpha < -0.5$
may not seem to be significantly different numerically from the
BA value  $\gamma = 3.0$ to claim  that  
it belongs to a different 
universality class. However, as we will discuss later,
the behaviour of the clustering coefficients are also significantly different 
here, which will support this claim.
All the above results were obtained for a network with $N = 20000$ and
using 100 different realisations of the network.

 We calculated the average shortest path lengths and diameter
 of the model at the phase boundary and found that the these
 two indeed scale logarithmically with the number of nodes in the network
 at the phase boundary indicating that the scale-free network
 also has a small diameter.

The clustering properties of this model
are studied in detail in an attempt to compare the results with that of the real networks.
In order to study clustering we kept $m= m_0 = 3$.
 Defining the exponents $a$ and $b$ in the following way  
 \begin{equation}
 {\cal C}(N) \propto N^{-a}
 \end{equation}
 and
\begin{equation}
{\cal C}(k) \propto k^{-b},
\end{equation}
  we find that  $a$ and $b$  depend on the values of $\alpha$ and $\beta$.
 Fig. 2 shows the behaviour of the clustering coefficients 
 ${\cal C}(N)$ 
 on the critical curve of the phase diagram
 as a function of the number of nodes.
 We find that for $\alpha > -0.5$, the data is consistent with 
 the behaviour ${\cal C}(N) \propto N^{-0.75}$. 
The slope of the curves decrease as we go away from $\alpha = -0.5$ to
higher negative values indicating that $a$ increases.
This is consistent with the idea that as $\alpha$ is made more negative,
the nodes get connected to the nearer ones making the clustering tendency higher.
A curious feature of  ${\cal C}(N) $
is that it actually increases with $N$ for large negative values of $\alpha$,
e.g., at the critical point corresponding to  $\alpha = -4.0$, $a$ becomes negative.
However, as the maximum value of ${\cal C}(N) $ can be unity, we believe that
a negative value of $a$ indicated that ${\cal C}(N)$ converges to 
a finite value for $\cal(N) \to \infty$ 
for large values of $\alpha$ on the negative side.

\begin{center}
\begin{figure}
\includegraphics[clip, width= 5cm]{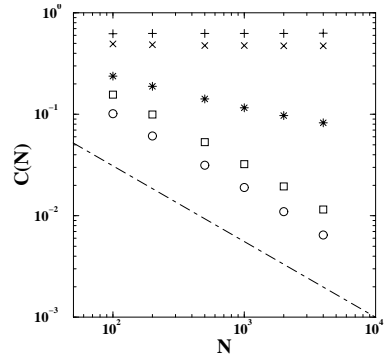}
\caption{The clustering coefficients as a function of $N$, the number of
nodes 
for different values of $\alpha$ on the phase boundary ($\alpha 
= -3.0, -2.0, -1.0, 0 $ and $1.0$ from top to bottom). 
The gradient in the log-log plot gives the value
of $a$.}
\end{figure}
\end{center}

Although the scaling behaviour of  ${\cal C}(N) $ remains same for all 
$\alpha > -0.5$, calculation of  ${\cal C}(N) $ for a fixed 
$N$ shows that on   increasing $\alpha$ the clustering decreases, a result
one can intuitively guess as for large positive $\alpha$, the
nodes get connected to nodes at large distances making
the clustering tendency lesser.

Fig. 3  shows the variation of ${\cal C}(k)$ against $k$ on the
phase boundary. ${\cal C}(k)$ is  more or less a constant for $\alpha > -0.5$,
but for larger negative values of $\alpha$  shows a decrease with $k$.
The behaviour of ${\cal C}(k)$ shows a  clear power law decay
for very large negative values of $\alpha$. 
This is a feature found in most real-world networks.

\begin{center}
\begin{figure}
\includegraphics[clip, width= 5cm]{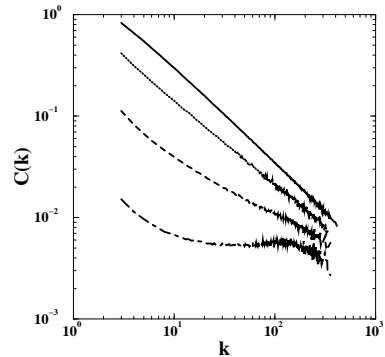}
\caption{The clustering coefficients as a function of $k$, the number of
nodes for different values of $\alpha$ on the phase boundary ($\alpha 
= -3.0, 1,5, -1.0$ and $0$ from top to bottom). The gradient in the log-log plot gives the value
of $b$.}
\end{figure}
\end{center}

We plot the values of $a$ and $b$  in Fig. 4 at  the critical points ($\alpha_c, \beta_c$) 
as a function of $\alpha$ as we are more interested in the role of the 
spatial distance dependence 
of   the network.
For $\alpha = 0$ and $\beta = 1$, we get the known values
$a = 0.75$ and $b = 0$. For all values of $\alpha > -0.5$,
the values of $a$ and $b$ remain the same on the critical 
phase boundary ($\beta_c = 1$) and are the equal to that  of
the BA model. For $\alpha < -0.5$, the values of $a$ and $b$ 
are different at different  points of the   phase boundary. In fact,
the value of $a$ decreases  while  $b$ increases towards 1 as $\alpha$
approaches higher negative values.

We have also studied the behaviour of the clustering coeffcients
in the regions of the phase diagram where it is not scale-free. In the region
where there is gelation, ${\cal C} (k)$ shows a power law behaviour again.
 This is expected as most of the nodes get attached to a single node
and the clustering coefficient decreases as a result. In the  region where stretched
exponential behaviour is observed, the clustering coefficient does not
show reasonable dependence on $k$ at large values of $k$. 

%
%

\begin{center}
\begin{figure}
\includegraphics[clip, width= 5cm]{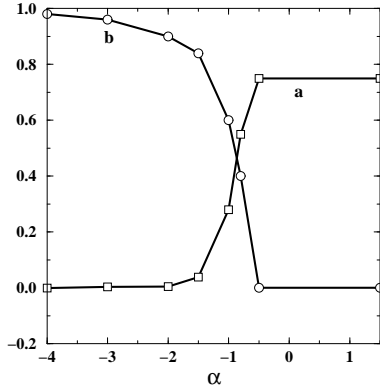}
\caption{The values of $a$ and $b$  on the phase boundary ($\alpha = \alpha_c,
\beta = \beta_c$) as a function of $\alpha$.
}
\end{figure}
\end{center}

In the present model,    ${\cal C} (k) \propto k^{-b}$ with a non-zero value
of $b$ for
 $\alpha < -0.5$ (with  $\beta = \beta_c$) for which
 the network is scale-free and also has a small diameter. 
 The power law behaviour  of ${\cal C} (k) $
 is obtained as a natural consequence of (1) without adding further
 steps in the growth process as in the other models considered in recent
 literature.
Surprisingly,
 both the present model and some of the other models considered 
 earlier \cite{RB,Doro,klemm,szabo} 
 give scale-free behaviour as well as 
${\cal C} (k) \propto k^{-b}$ (with $b \ne 0$) although they differ
by an important factor - the spatial dependence or geographical 
organisation. Hence, it is  not possible to guess whether
there is any such organisation  present  in the
network  simply by knowing $b$. The real networks with geographical
organisation in fact show that $b = 0$, a result we can obtain from the present 
model when the spatial dependence given by $\alpha$ becomes irrelevant
and it becomes equivalent to the BA model.
 Hence we conclude that geographical
organisation is not the key factor responsible for the result $b = 0$.
And the result $b \ne 0$ can be achieved even after  incorporating 
distance dependent factors.

Our present results are for a one dimensional network. But as observed in
\cite{mannaps}, 
when  $\alpha \ne 0$  and $\beta =1$, 
the two dimensional network
gives results which are qualitatively  similar
to those obtained in one dimension, we believe that in higher
dimensions also one would get similar results.

To summarise, we have studied a growing network in the Euclidean                             space
where the link attachment probability is controlled jointly by two
competing factors i.e., the preferential attachment and the
 magnitude of the link length. These two factors are tuned by the
 parameters $\alpha$ and $\beta$ as defined in Eqn. (1). A  critical
 boundary in the $\alpha-\beta$ phase plane separates the network
  from its ``gel'' phase to the ``stretched exponential'' phase. However on
  the boundary between the two phases the network is scale-free.
   Numerical simulations on a one dimensional system indicates that on
   the critical boundary the network crosses over from a BA universality
      class ($\alpha > -0.5$) to a new universal scale free behaviour (($\alpha
           < -0.5$). The calculation of the 
	   exponents $a$ and $b$ for the  clustering
	   coeffcients defined in  Eqns (2) and (3) show that their  values 
	    are non-universal 
in the region  $\alpha < -0.5$ on the phase boundary, 
with an indication that $a$ converges to zero and $b$ converges
to unity 
as $\alpha$ approaches large negative values. 
Thus  the network can be tuned to  have 
 different   clustering properties on the phase boundary.

E-mail: parongama@vsnl.net,paro@cubmb.ernet.in; manna@boson.bose.res.in.

Acknowledgements: PS acknowledges support form DST grant SP/S2/M-11/99.

\end{multicols}

\end{document}